\documentclass[prd,aps,preprint,showpacs]{revtex4}

\usepackage{graphicx}
\usepackage{color}
\usepackage{amsfonts}
\usepackage{amsmath}

\def\ul#1{\underline{#1}}
\def\hpsi{\hat{\Psi}}
\def\fn{f^{(n)}}
\def\fan{{f_A^{(n)}}}
\def\psn{\psi^{(n)}}

\def\n{{(n)}}

\newcommand{\Z}     {\mathbb{Z}_2}
\newcommand{\ZZ}     {\Z\times\Z'}
\newcommand{\SZZ}     {S^1/\Z\times\Z'}
\newcommand{\ai}     { a_{_{\rm IR}}}
\newcommand{\au}     { a_{_{\rm UV}}}
\newcommand{\zi}     { z_{_{\rm IR}}}
\newcommand{\zu}     { z_{_{\rm UV}}}
\newcommand{\man}     {{ m_A^{(n)} }}
\newcommand{\xan}     {{ x_A^{(n)} }}
\newcommand{\beq}     {\begin{equation}}
\newcommand{\eeq}     {\end{equation}}
\newcommand{\bea}     {\begin{eqnarray}}
\newcommand{\eea}     {\end{eqnarray}}

\newcommand{\hPsi}     {\widehat{\Psi}}
\newcommand{\es}     { \epsilon}
\newcommand{\ves}     { \varepsilon}
\newcommand{\lm}     { \lambda_v}
\newcommand{\gm}     { \gamma}
\newcommand{\no}     { \nonumber}

\newcommand{\Wsh}     {{ \rlap/W }}
\newcommand{\gtb}     {{ g_{_{Wtb}} }}

\begin{document}


\preprint{ KIAS--P04049 \hspace{1ex} hep-ph/041XXXXX }

\title{Kaluza-Klein masses of bulk fields \\
with general boundary conditions in AdS$_5$ }

\author{Sanghyeon Chang\footnote{schang@phya.yonsei.ac.kr}}
\affiliation{Department of Physics and
IPAP, Yonsei University, Seoul 120-749, Korea}

\author{Seong Chan Park\footnote{spark@kias.re.kr}}
\affiliation{ Korea Institute for Advanced Study, 207-43,
Seoul 130-722, Korea  }

\author{
Jeonghyeon Song\footnote{jhsong@konkuk.ac.kr}}
\affiliation{Department of Physics, Konkuk University,
                   Seoul 143-701, Korea}

\begin{abstract}
Recently bulk Randall-Sundrum theories with the gauge group $SU(2)_L
\times SU(2)_R \times U(1)_{B-L}$ have drawn a lot of interest as an
alternative to electroweak symmetry breaking mechanism. These models
are in better agreement with electroweak precision data since
custodial isospin symmetry on the IR brane is protected by the
extended bulk gauge symmetry. We comprehensively study, in the
$S^1/\ZZ$ orbifold, the bulk gauge and fermion fields with the
general boundary conditions as well as the bulk and localized mass
terms. Master equations to determine the Kaluza-Klein (KK) mass
spectra are derived without any approximation, which is an important
basic step for various phenomenologies at high energy colliders. The
correspondence between orbifold boundary conditions and localized
mass terms is demonstrated not only in the gauge sector but also in
the fermion sector. As the localized mass increases, the first KK
fermion mass is shown to decrease while the first KK gauge boson
mass to increase. The degree of gauge coupling universality
violation is computed to be small in most parameter space, and its
correlation with the mass difference between the top quark and light
quark KK mode is also studied.

\end{abstract}

\maketitle
\thispagestyle{empty}

\section{Introduction}

The origin of electroweak symmetry breaking (EWSB) has still
remained to be explored by experiments. In the standard model (SM),
EWSB occurs spontaneously as the Higgs field develops vacuum
expectation value (VEV). This Higgs mechanism is, however, regarded
unsatisfactory since the Higgs potential is introduced just for the
purpose of EWSB itself. Furthermore it is extremely unstable against
radiative corrections and thus UV physics, creating the so-called
gauge hierarchy problem. Most of models for new physics beyond the
SM pursue more natural EWSB mechanism. According to symmetry
breaking coupling strength, new models are divided into two classes:
One is a weakly coupled theory with a high cut-off scale
and the other is a strongly coupled theory~\cite{dynamical}.

Recently it is shown that these two different classes can be
related by AdS/CFT
duality~\cite{Maldacena:1997re,Gubser:1998bc,Witten:1998qj,Aharony:1999ti}:
A four-dimensional (4D) theory with a \emph{strongly} coupled
sector conformal from the Planck scale to the weak scale is dual
to a 5D \emph{weakly} coupled Randall-Sundrum model-1
(RS1)~\cite{Randall:1999ee,Randall:1999vf,Arkani-Hamed:2000ds,Rattazzi:2000hs,Perez-Victoria:2001pa}.
The RS1 model has one extra spatial dimension of a truncated
AdS space, the orbifold of $\SZZ$ without
the assumption of periodic boundary condition.
The fixed point under $\Z$ parity transformation
is called the UV brane
and that under $\Z'$ parity the IR brane.
In the original RS
scenario, all the SM fields are
confined on the TeV brane~\cite{Randall:1999ee}.
Since a localized field in the 5D
theory is dual to the TeV-scale composite in the strong sector
of the 4D theory, the phenomenological aspects of a localized field depend
sensitively on the unknown UV physics.
This feature aroused great interest in bulk RS
theories~\cite{Davoudiasl:1999tf,Chang:1999nh,Kim:2002kk,Grossman:1999ra,
Gherghetta:2000qt, Huber:2000fh, Huber:2000ie}.
As weakly coupled effective field theories, their phenomenological
implications become more reliable.
For example it is feasible to discuss the RG running of
gauge couplings and their
unification~\cite{Pomarol:2000hp,Goldberger:2002hb,Agashe:2002bx,
Contino:2002kc,Randall:2001gb,Randall:2001gc,Choi:2002ps,Agashe:2002pr,
Goldberger:2002pc,KChoi}.
To solve the gauge hierarchy problem, however,
the Higgs field should be localized
on the IR-brane.

A naive extension of the RS1 model by releasing the SM gauge and
fermion fields in the bulk, however, has troubles with electroweak
precision data, particularly with the Peskin-Takeuchi $T$
parameter~\cite{Csaki:2002gy, Hewett:2002fe,Burdman:2002gr}. This
problem is attributed to the lack of isospin custodial symmetry.
Recently a bulk gauge symmetry of $SU(2)_R$ has been added,
which is used to restore a gauge version of custodial symmetry in the
bulk~\cite{Agashe:2003zs, Burdman:2004rz}. Another rather radical
solution to EWSB in this framework is the Higgsless theory: The
gauge symmetry breaking is due to non-trivial orbifold boundary conditions.
Non-zero SM gauge boson masses are nothing
but the first Kaluza-Klein (KK) mode
mass~\cite{Csaki:2003dt,Csaki:2003sh,Csaki:2003zu}. From the AdS/CFT
correspondence, we can interpret this model as a dual of a
technicolor model.

Both models with~\cite{Agashe:2003zs} and
without~\cite{Csaki:2003dt} a Higgs boson incorporate two kinds of
new ingredients in the phenomenological view point. First, we have
new gauge fields $\overrightarrow{W}^\mu_{R}$ of $SU(2)_R$,
introduced for the custodial isospin symmetry. At high energy
colliders, they appear as KK excitations with TeV scale masses since
the $\ZZ$ parity of $\overrightarrow{W}^\mu_{R}$ is not $(++)$.
Secondly, new bulk fermions are also required.
The $SU(2)_R$ symmetry,
which promotes the SM
right-handed fermions to the doublets,
is broken by the UV orbifold boundary conditions:
in the Agashe-Delgado-May-Sundrum (ADMS) model~\cite{Agashe:2003zs},
for example, ${W}^\pm_{R}$ fields
have definite $(-+)$ parity; the SM right-handed up quark
with $(++)$ parity should couple with a new right-handed down-type
quark with $(-+)$ parity for $\ZZ$ invariant action.

Since the bulk RS models with custodial isospin symmetry are
compatible with electroweak precision data, its
phenomenological probe should await experiments at future
colliders~\cite{Barbieri:2004qk,Barbieri:2003pr,
Nomura:2003du,Foadi:2003xa,Davoudiasl:2003me,
Davoudiasl:2004pw,Burdman:2003ya,Cacciapaglia:2004jz,Cacciapaglia:2004rb,
Birkedal:2004au}.
Exact KK mass spectra of the bulk gauge boson and fermion
are of great significance.
In the ADMS model where
the $\ZZ$ parity is conserved at tree level,
for example,
the decay of new gauge bosons with $(-+)$
parity into the SM particles with $(++)$ parity can be limited and thus
long-lived.
We will derive, without any approximation, master equations for the KK masses
particularly with the general localized mass terms.
Special focus is on the KK masses of the top quark, on which the effect of
the localized Yukawa coupling is significant.
Contrary to the gauge bosons case,
the first KK mode mass of top quark decreases with increasing top Yukawa coupling.
We will also suggest a phenomenologically dramatic case,
called the KK mode degenerate case,
where the first and second KK masses of fermions are degenerate
without the localized Yukawa coupling.
Another interesting feature of this case is
that the mass drop of the first KK mode
by the localized Yukawa coupling is maximized.
It is very feasible, therefore,
that the first signal of KK fermion comes from
the top quark mode.

It is also worthwhile to distinguish
the role of UV-brane localized VEV (parameterized by a dimensionless parameter $\au$)
and that of IR-brane localized VEV (parameterized by $\ai$)
in the generation of  the
zero-mode mass of a gauge boson with (++) parity.
Generically either $\au$ or $\ai$
generates TeV scale mass for the zero mode,
which would vanish with $\au=\ai=0$.
We will show that quite different is the way
to generate the zero mode mass:
$\ai$
gradually increases the zero-mode mass, while even small $\au$ (but
larger than about $10^{-15}$) lifts up the zero-mode
mass to TeV scale at one stroke.

Another interesting issue is the theoretical relation
between the Higgsless model and
the ADMS model.
This correspondence in the gauge sector was
pointed out in Ref.~\cite{Nomura:2003du}.
Similar correspondence in the bulk fermion sector is deserved to study also.
Based on
the exact formulae of KK masses,
we will show that
the bulk fermion field with non-trivial orbifold boundary conditions
in the Higgsless theory
can be
understood through the VEV of a localized scalar field.
This will complete the understanding of
orbifold boundary conditions.

Inevitable deviation of gauge coupling unification,
denoted by $\delta g_{Wtb}$,
shall be studied.
We will focus on its correlation with the top
quark mass spectra.
If this correlation is strong enough,
it can be a valuable information
since the magnitude of $\delta g_{Wtb}$ in most parameter space
is too small to be probed at hadron colliders.
Restricted to the KK mode degenerate case,
we will show a significant correlation between the $\delta g$
and the top quark KK mode mass relative to light quark KK mode mass.

This paper is organized as follows. In
Sec.~\ref{sec:bgauge}, we give the general setup for
the bulk gauge boson.
By solving wave equations with the
brane localized mass terms, we derive the master equations for the KK masses.
The interpolation between the
ADMS model and the Higgsless model are to be
understood as a consequence of master equations in a limiting
case. Some numerical values of KK states are also
presented. Section \ref{sec:bfermion} deals with the KK masses
of a bulk fermion without the brane localized
mass term.
More delicate case with the brane
localized Yukawa coupling is considered in
Sec.~\ref{sec:bfermionYukawa}.
The possibility of gauge coupling universality violation is
also studied in Sec.~\ref{sec:universality}, which arises due to
the deviation of KK zero mode functions by brane localized
masses. In Sec. \ref{sec:conclusion},
we present the summary and conclusions.

\section{Bulk gauge bosons}
\label{sec:bgauge}

\begin{figure}[tbh]
  \includegraphics[scale=1]{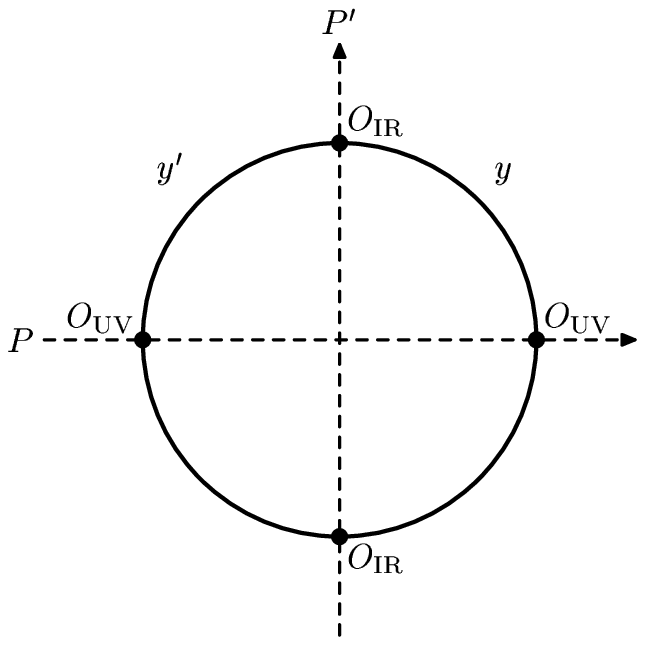}
  \caption{\label{fig:S1ZZ}
  The $S^1/\Z\times\Z'$ orbifold.}
\end{figure}

We consider a gauge theory in a five-dimensional warped spacetime
with the metric given by
\begin{equation}
\label{eq:RSmetricy}
ds_5^2 \equiv g_{MN}dx^M dx^N= e^{-2\sigma(y)}(dt^2-d\vec{x}\,^2) - dy^2,
\end{equation}
where $y$ is the fifth dimension coordinate and $\sigma(y) = k|y|$.
The theory is to be compactified on the
$S^1/\Z \times \Z'$ orbifold,
which is a circle of radius $r_c$
with two reflection symmetries
under $\Z:y\to -y$ and $\Z':y'\to -y'$
$(y'=y-\pi r_c/2)$ as depicted in Fig.~\ref{fig:S1ZZ}.
Often the conformal
coordinate of $z\equiv e^\sigma/k$ is useful:
\begin{eqnarray}
\label{eq:RSmetricy2}
ds_5^2=\frac{1}{(k z)^2}\left(dt^2-d\vec{x}\,^2-dz^2\right).
\end{eqnarray}
Since $y$ is confined
in $0\leq y \leq L ~(L\equiv \pi r_c/2)$,
$z$ is also bounded
in $1/k \leq z \leq 1/T$.
Here $T$ is the effective electroweak scale,
defined by $T \equiv e^{-kL} k \equiv \es\, k$.
With $k L \approx 35$,
the warp factor $\es(\equiv e^{-kL})$
reduces
$T$ at TeV scale from $k$ at Planck scale:
With this scaling the gauge hierarchy problem is answered.
The space of $S^1/\Z\times \Z'$
accommodates two fixed points,
the $\Z$ fixed point at $\zu=1/k$
(called the UV brane) and
the $\Z'$ fixed point at
$\zi=1/T$ (called the IR brane).

The action for a 5D $U(1)$
gauge field is
\begin{eqnarray}
\label{eq:S-A}
S_{\mathrm{gauge}} = \int d^4 x dz
\sqrt{G}\left[-\frac{1}{4}g^{MP}g^{NQ}F_{MN}F_{PQ}
+\frac{1}{2}M^2 g^{MN}A_M A_N\right],
\end{eqnarray}
where $G$ is the determinant of the AdS metric, $F_{M
N}=\partial_M A_N- \partial_N A_M$.
The general mass term $M^2(z)$,
including the case where the gauge symmetry is broken in the bulk, is
\begin{eqnarray}
M^2(z)=\au^2 k \,\delta(z-\zu)+\ai^2 k \,\delta(z-\zi)+ b^2 k^2,
\end{eqnarray}
where the dimensionless
$b$ and $\au (\ai)$ parameterize
the bulk mass and
the localized mass on the UV (IR)
brane, respectively.
Note that $b$ breaks the gauge symmetry.
%

The KK expansion of the dimension 3/2 field $A^M(x,z)$ is
\begin{eqnarray}
\label{eq:KKexpA}
A_{\nu} (x,z) = \sqrt{k} \sum_n A_\nu^{(n)}(x) \fan (z),
\end{eqnarray}
where the mode function
$\fan(z)$ is dimensionless.
With the following equation of motion for $\fan(z)$
\begin{eqnarray}
\label{eq:eqfA}
-z \partial_z \left(\frac{1}{z}\partial_z
\fan(z)\right)+\frac{M^2(z)}{k^2 z^2}\fan(z)=\man^2 \fan(z),
\end{eqnarray}
and the normalization of
\beq
\label{eq:normA}
\int \frac{d z }{z} f_A^{(n)} f_A^{(m)} = \delta_{nm}
\,,
\eeq
the action in Eq.~(\ref{eq:S-A}) describes
a tower of massive KK gauge bosons:
\beq
S_{\mathrm{gauge}}
= \sum_n \int d^4 x
\left[
- \frac{1}{4}\eta^{\mu\rho}\eta^{\nu\kappa}
F^{(n)}_{\mu\nu}F^\n_{\rho\kappa}
- m_A^\n \eta^{\mu\nu}A^{\n}_{ \mu} A^\n_\nu
\right]
\,,
\eeq
where $F^\n_{\mu\nu}=\partial_\mu A^\n_\nu -\partial_\nu A^\n_\mu$.
The general solution of Eq.~(\ref{eq:eqfA})
in the bulk ($\zu < z <\zi$) is
\begin{eqnarray}
\label{eq:fAn}
\fan(z)=  \frac{z}{N_A^\n} \left[ J_\nu(\man z)
+ \beta_A^\n Y_\nu(\man z)
\right],
\end{eqnarray}
where $\nu = \sqrt{1+b^2}$ and $N_A^\n$ is
determined by the normalization condition in Eq.~(\ref{eq:normA}).

Boundary conditions on the two branes
specify the constant $\beta_A^\n$.
If the mode function
$f_A^\n(z)$ has $\Z$- or $\Z'$-even parity,
Neumann boundary condition applies as
\begin{eqnarray}
\label{eq:NeumannBC}
\left.
\frac{d \fan}{dz}
\right|_{z=z_i} = (-1)^{P_i} \frac{a_i^2}{2} \fn|_{z=z_i},
\quad
i={\mathrm{UV},~\mathrm{IR}}
\end{eqnarray}
where $P_{\rm UV}=2$, and $P_{\rm IR}=1$.
The sign difference between the UV and
IR brane is due to the directionality of the derivative at the
boundary points.
Here physics is essentially
similar to the case where the electric field near the
conducting boundary is determined by the charge localized on the
conducting plane.
For
$\Z$- or $\Z'$-even function,
the $\beta_A^\n$ coefficient is 
%
\begin{eqnarray}
\label{eq:RN}
-\beta_A^{(n)} \mid _{\rm even}&=& \frac{\left(-(-1)^{P_i}
\frac{a_i^2}{2}+1-\nu \right)J_\nu(\man z_i)+\man z_i J_{\nu-1}(\man
z_i)}{\left(-(-1)^{P_i} \frac{a_i^2}{2}+1-\nu \right)Y_\nu(\man
z_i)+\man z_i Y_{\nu-1}(\man z_i)} \\ \nonumber
&\equiv&
R_N (a_i, \nu,\man z_i).
\end{eqnarray}
If the function $\fan$
has $\Z$- or $\Z'$-odd parity at the corresponding boundary,
the Dirichlet
boundary condition applies as
\begin{eqnarray}
\label{eq:DirichletBC}
\left. \fan \right|_{z=z_i} =0\,.
\end{eqnarray}
Their $\beta_A^\n$'s are
\beq
\label{eq:RD}
-\left. \beta_A^{(n)} \right|_{\rm odd}
= \frac{J_\nu (\man z_i)}{Y_\nu
(\man z_i)}
\equiv R_D( \nu,\man z_i).
\eeq
Note that the Dirichlet boundary condition is
independent of the localized mass term $a_{_{\rm UV,IR}}$.

In the $S^1/\Z\times\Z'$ orbifold,
four different $\Z\times\Z'$ parities are possible as
$(++),(+-),(-+)$ and $(--)$.
Since two boundary conditions
doubly constrain a single \emph{constant} $\beta_A^\n$ coefficient,
the KK mass $\man$ is determined
by the following master equations:
\bea
\label{eq:pp}
\bullet ~~(++) :\quad
R_N(\au,\nu,\man \zu) &=& R_N (\ai, \nu, \man \zi),
\\ \label{eq:pm}
\bullet ~~ (+-) : \quad
R_N (\au,\nu, \man \zu) &=& R_D( \nu,\man \zi),
\\ \label{eq:mp}
\bullet ~~ (-+) :\quad \hspace{24pt}
R_D (\nu, \man \zu) &=& R_N(\ai, \nu, \man \zi ),
\\ \label{eq:mm}
\bullet ~~ (--) :\quad \hspace{24pt}
R_D (\nu, \man \zu) &=& R_D ( \nu, \man \zi)
\,.
\eea

\begin{figure}[tb]
    \includegraphics[scale=1.0]{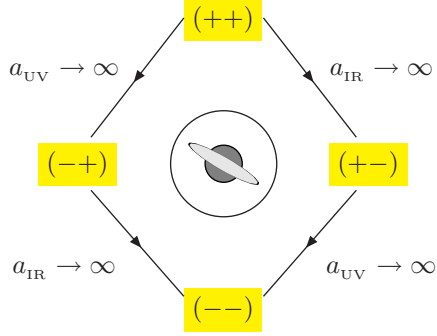}
    \caption{Diagram shows relationship between
different boundary conditions by dialing vacuum expectation value
of localized Higgs fields. This is the underlying physics in the
interpolation between the theories of gauge symmetry breaking by a
localized Higgs field and by a technicolor-like strong dynamics in
the AdS dual picture.}
    \label{fig:diagram}
\end{figure}

From the functional forms of $R_N$ and $R_D$,
the $\Z\times \Z'$ parity
can be understood through the effect of large localized Higgs VEV.
In the
large $a_i$ limit, $R_N$ approaches $R_D$:
\begin{eqnarray} \label{large vev limit}
\lim_{a_i \rightarrow \infty}R_N (a_i,\nu, \man z_i)
=\lim_{a_i \rightarrow \infty}
\frac{-(-1)^P \frac{a_i^2/}{2} J_\nu (\man z_i)}
{-(-1)^P \frac{a_i^2/}{2} Y_\nu (\man
z_i)}= R_D (\nu,\man z_i).
\end{eqnarray}
This happens because the 5D wave function is expelled
by large VEV of the localized Higgs field.
In the large $a_i$ limit, therefore,
the KK masses
of the $\Z$- or $\Z'$-even gauge field
become identical with those of $\Z$- or $\Z'$-odd field.
For example, the KK masses of
a $(++)$ gauge field with large $\au$
are the same as that of a $(-+)$ field without localized mass.
Usually this behavior is expressed that
the $(++)$ gauge field
mimics  $(-+)$ field.
Similarly, the
$(++)$ gauge field with large $\ai$
mimics the $(+-)$ field;
the $(++)$ field with both large $\au$ and $\ai$
mimics the $(--)$ field.
Figure \ref{fig:diagram} summarizes all the correspondences.
These
relations could be interpreted as the origin of the interpolation
between the ADMS model and the Higgsless model in the
AdS dual picture.

\begin{figure}
\includegraphics{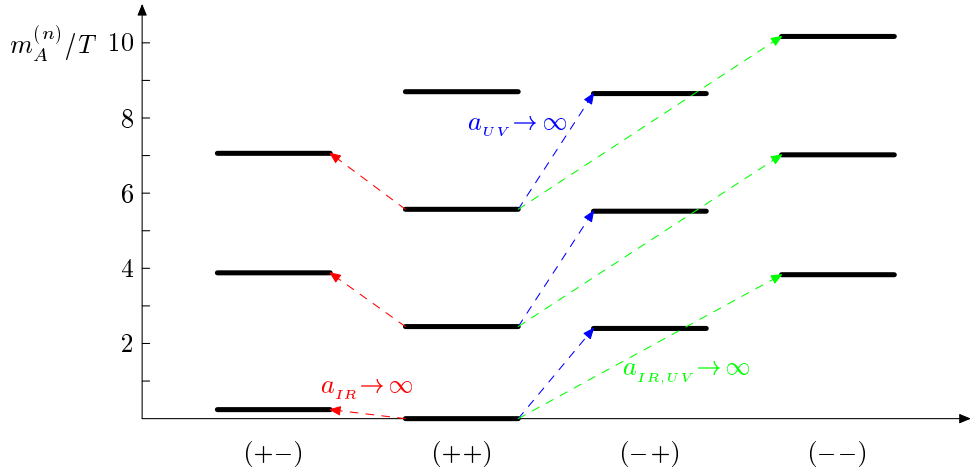}
\caption{Kaluza-Klein
masses of a bulk gauge boson in unit of $T$
when $b=\au=\ai=0$.}
\label{fig:bulkA}
\end{figure}

For the numerical calculation, we assume the bulk mass parameter $b$
to be zero.
In Fig.~\ref{fig:bulkA}, we present the KK masses of a bulk gauge boson
in unite of $T$
without any localized masses, i.e., $\au=\ai=0$.
The RS metric alone determines the KK mass spectra.
It is clear that only the bulk gauge boson with $(++)$ parity allows zero mode.
A remarkable feature
is the substantially light mass of
the first KK mode with $(+-)$ parity.
With $T \sim $TeV,  $m_{A(+-)}^{(1)}$
can be of order 100 GeV.
Since the $(+-)$ parity mode is equivalent to the
$(++)$ parity mode in the limit of large $\ai$,
this feature suggests the possibility of the gauge symmetry breaking
by orbifold boundary conditions without the Higgs mechanism.
Numerically we have
\bea
{m_{A(++)}^{(1)}}
&\approx &2.45\,{T},\quad
{m_{A(++)}^{(2)}}
\approx5.57\,{T},\quad
{m_{A(++)}^{(3)}}
\approx8.70\,{T},
\\ \no
{m_{A(+-)}^{(1)}}
&\approx &0.24\,{T},\quad
{m_{A(+-)}^{(2)}}
\approx 3.88\,{T},\quad
{m_{A(+-)}^{(3)}}
\approx7.06\,{T},
\\ \no
{m_{A(-+)}^{(1)}}
&\approx &2.40\,{T},\quad
{m_{A(-+)}^{(2)}}
\approx 5.52\,{T},\quad
{m_{A(-+)}^{(3)}}
\approx 8.65\,{T},
\\ \no
{m_{A(--)}^{(1)}}
&\approx &3.83\,{T},\quad
{m_{A(--)}^{(2)}}
\approx 7.02\,{T},\quad
{m_{A(--)}^{(3)}}
\approx 10.17\,{T}.
\eea

\begin{figure}[bth]
    \includegraphics[scale=1.0]{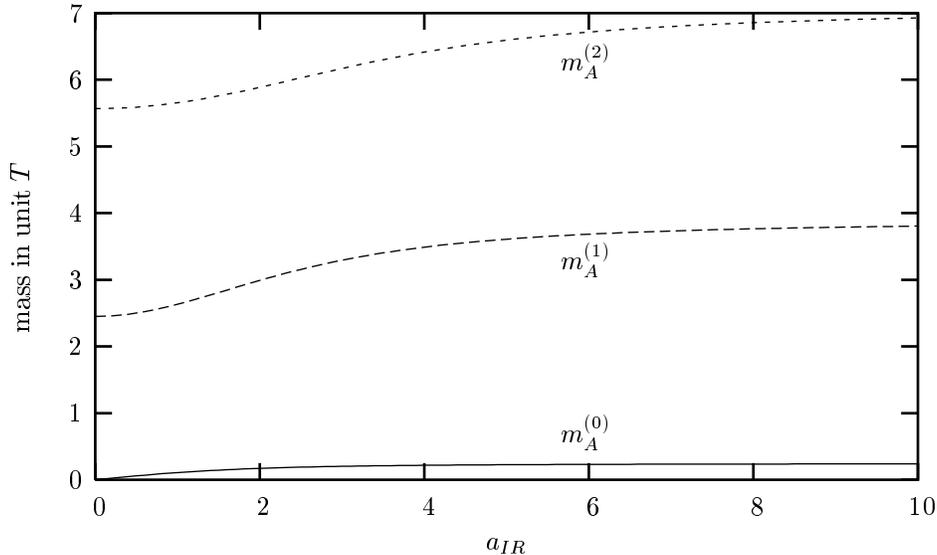}
    \caption {The lowest a few KK
masses of a bulk gauge boson with
$(++)$ parity for varying $a_{\rm IR}$. At large $a_{\rm IR}$ limit, we can
easily see the masses are saturated.
    }
    \label{fig:bulkAair}
\end{figure}

If the SM gauge symmetry of $SU(2)_L \times U(1)_Y$
is spontaneously broken by the localized Higgs VEV,
the value of $a_{\rm IR}$ becomes non-zero. Figure \ref{fig:bulkAair}
shows, as functions of $\ai$,
a few lowest KK masses of a bulk gauge boson with
$(++)$ parity.
In the small $a_{\rm IR}$ limit, the zero mode
KK mass increases gradually with $\ai$.
As
$a_{\rm IR}$ becomes large, the rise of KK masses is saturated,
eventually into the KK masses of $(+-)$ parity modes.

When the UV brane mass parameter $\au$ turns on, however,
the rise of zero mode mass with the $\au$ is \emph{not}
gradual even in the small $\au$ limit.
To be specific, let us focus on the master equation for
the $(++)$ gauge field in Eq.~(\ref{eq:pp})
with $\ai=0$.
Denoting the KK mass in unit of $T$ by
$x_A^\n \equiv \man/T$,
$x_A^\n$ is the solution of
\beq
\label{eq:au}
\frac{J_0(\xan)}{Y_0(\xan)}=
\frac{-\frac{\au^2}{2}J_1(\es \xan)+\es \xan J_0(\es \xan)}
{-\frac{\au^2}{2}Y_1(\es \xan)+\es \xan Y_0(\es \xan)}
\,,
\eeq
whose the right-handed side in the limit of $\es \ll 1$ is
$
\left(
-{1}/{4 }+{1}/{\au^2}
\right)\es^2 \,\xan^2
$.
In order to avoid another hierarchy in the theory,
even small $\au$ is assumed larger than $\es=T/k \sim 10^{-15}$.
Then
the right handed side of Eq.~(\ref{eq:au}) is technically zero,
and the $\au$-dependence disappears.
As soon as the $\au$ above $10^{-15}$ turns on,
the KK masses of the $(++)$ gauge field jump into those of the $(-+)$ field.
In summary, the SM gauge bosons mass in the AdS$_5$ background
can be generated either by orbifold boundary conditions
or by the IR-brane localized Higgs VEV.

\section{Bulk fermion field without the localized mass}
\label{sec:bfermion}

In 5D spacetime,
the Dirac spinor is the smallest irreducible representation
of the Lorentz group.
Its 5D action is
\begin{eqnarray}
S_{\rm fermion}=\int d^4x dy \sqrt{G}
\left[
\,\frac{i}{2}
\overline{\Psi}\,\Gamma^{\ul{A}}\,e_{\ul{A}}{}^A
\partial_A\Psi -\frac{i}{2}(\partial_A \overline{\Psi})\,\Gamma^{\ul{A}}\,e_{\ul{A}}{}^A
 + m_D \overline{\Psi}\Psi \right]
 \,,
\end{eqnarray}
where
$e_{\ul{A}}{}^A=
\mbox{diag}(e^\sigma,e^\sigma,e^\sigma,e^\sigma,1)$ is the inverse
f\"{u}nfbein,
$\Gamma^{\ul{M}}=\left(\gamma^\mu,\,i\gamma_5\right)$,
$\sqrt{G}=e^{-4\sigma}$,
and $
\{\Gamma_{\ul{M}},\,\Gamma_{\ul{N}}\}
=2\eta_{\ul{M}\,\ul{N}}=2{\rm diag}\left(+,-,-,-,-\right)$.
In order to make good use of the $\Z\times\Z'$ parity,
we employ the extra dimensional coordinate $y$
in Eq.~(\ref{eq:RSmetricy}).
With the redefinition of
$\hPsi\equiv
e^{-2\sigma}\Psi $ and the relation of
$\partial_y \Psi =e^{2\sigma}(2 \sigma' + \partial_y )\hPsi
$,
the action can be simply written by
\begin{eqnarray}
S_{\rm fermion}=\int d^4x \,dy
\left[ \,\overline{\hPsi} e^\sigma
i\gamma^\mu\partial_\mu \hPsi -\frac{1}{2}
\overline{\hPsi} \gamma_5\partial_y \hPsi
+\frac{1}{2}(\partial_y \overline{\hPsi})\gamma_5 \hPsi
 + m_D \overline{\hPsi}\hPsi \right]\,.
\end{eqnarray}

Under the $\Z$ orbifold symmetry,
a bulk fermion has two possible transformations of
\beq
\label{eq:Z2sym}
\gamma_5 \Psi_\pm (x,-y)\,={}\pm\Psi_\pm (x,y)
\,.
\eeq
To understand this $\Z$ symmetry more easily, we
decompose the bulk fermion
in terms of KK chiral fermions:
\beq
\hPsi(x,y)
\equiv \hPsi_L+\hPsi_R= \sqrt{k} \sum_{n}
\left[
\psi_L^\n(z) f_L^\n(y) + \psi_R^\n(z) f_R^\n(y)
\right]
\,.
\eeq
When $\Psi(x,y)$ is even under $\Z$, for example,
$f_L^\n$ is odd while $f_R^\n$ is even:
\begin{equation}
f_L^{(n)}(-y)=-f_L^{(n)}(y),\    f_R^{(n)}(-y) = f_R^{(n)}(y).
\end{equation}
Similar arguments for $\Z'$ symmetry can be made.
Therefore, the $\Z\times\Z'$ parity of $\Psi_L$ is always
opposite to that of $\Psi_R$.

Another tricky problem arises
when dealing with a bulk fermion in a finite interval.
To confirm the variational principle,
we separate the action into the bulk term ($S_B$) and the boundary
term ($S_{\partial B}$):
\begin{eqnarray}
\label{eq:SB}
S_B\!&=&\!\int\! d^4x dy \left[\overline{\hPsi}_L e^\sigma
i\gamma^\mu\partial_\mu \hPsi_L + \overline{\hPsi}_R
e^\sigma i\gamma^\mu\partial_\mu \hat{\Psi}_R
 + m_D (\overline{\hPsi}_L\hPsi_R +
 \overline{\hPsi}_R\hPsi_L)
 \right.
 \\ \no &&
 \left.
 -
\overline{\hPsi}_L \partial_y \hPsi_R
+\overline{\hPsi}_R \partial_y \hPsi_L\right],
\\
\label{eq:SrdB}
S_{\partial B} &=&
\int d^4x \,\frac{1}{2}
\left[\,\overline{\hPsi}_L\hPsi_R
-\overline{\hPsi}_R\hPsi_L \right]^{L}_0 \,
.
\end{eqnarray}
Since Dirac mass term in Eq.~(\ref{eq:SB}) is $\ZZ$-odd,
we define $m_D = c\,
\sigma'(y)=c\, k \mathrm{sign}(y)$.
Considering both boundaries,
$\sigma(y)$ is a periodic triangle wave function
and thus $\mathrm{sign}(y)$
is a periodic square wave function.

With the normalization of
\begin{equation}
\label{eq:normf}
\delta_{mn} =k \int^{L}_{0} dy\, e^\sigma f_L^{(n)}f_L^{(m)}
=k \int^{L}_{0} dy\, e^\sigma f_R^{(n)}f_R^{(m)} \,,
\end{equation}
and the equations of motion of
\begin{eqnarray}
\label{eq:eomffy}
 \partial_y  f_R^{(n)} - m_D  f_R^{(n)} & = & m^{(n)} e^\sigma  f_L^{(n)},  \nonumber\\
 -\partial_y  f_L^{(n)} - m_D  f_L^{(n)} & = & m^{(n)} e^\sigma  f_R^{(n)}
 \,,
\end{eqnarray}
the bulk action in Eq.~(\ref{eq:SB}) becomes the sum of KK fermion modes:
\begin{equation}
S_{\rm eff}
 = \int d^4 x \sum_n\left[\overline{\psi}^{(n)}_L i \gamma^\mu \partial_\mu
 \psi^{(n)}_L + \overline{\psi}^{(n)}_R i\gamma^{\mu} \partial_\mu
\psi^{(n)}_R - m^{(n)}(\overline{\psi}^{(n)}_L \psi^{(n)}_R +
\overline{\psi}^{(n)}_R \psi^{(n)}_L)
     \right] \ .
\label{eff0}
\end{equation}

The equations of motion
in the conformal coordinate $z= e^{\sigma(y)} /k$ are
\beq
\label{eq:eomffz}
\left(\partial_z + \frac{c}{z}\right) f_L^{(n)} = -m^{(n)}
f_R^{(n)},
\quad
\left(\partial_z - \frac{c}{z}\right) f_R^{(n)} = m^{(n)}
f_L^{(n)}
\,,
\eeq
which yield the general solutions of
\begin{eqnarray}
f_L^{(n)}(z) &=& \frac{\sqrt{z}}{N_L^{(n)}}
\left[J_{c+\frac{1}{2}}(m^{(n)} z) +
\beta_L^{(n)} Y_{c+\frac{1}{2}}(m^{(n)} z)\right], \nonumber \\
f_R^{(n)}(z) &=& \frac{\sqrt{z}}{N_R^{(n)}}
\left[J_{c-\frac{1}{2}}(m^{(n)} z) + \beta_R^{(n)}
Y_{c-\frac{1}{2}}(m^{(n)} z)\right]. \label{bessel}
\end{eqnarray}

Special properties of the Bessel function and the boundary condition
lead to the following simple relations:
\beq
\label{eq:betaN}
\beta_L^{(n)}=\beta_R^{(n)},\quad
N_L^{(n)}=-N_R^{(n)}\,.
\eeq
This is because either $f_L^\n$ or $f_R^\n$ is an continuous odd function
which vanishes at the boundary.
For example, consider the case where $f_R^\n$ is odd.
With Eq.~(\ref{eq:eomffz}),
we have
\begin{eqnarray}
\label{eq:oddzero}
&&\left. \fn_R \right|_{z=\frac{1}{k}}= 0,
\\
\label{eq:doddzero}
&& \left.\left( \partial_z + \frac{ c}{z} \right) \fn_L
\right|_{z=\frac{1}{k}}=0 .
\end{eqnarray}
Equation (\ref{eq:oddzero}) leads to
$\beta_R^{(n)}= -{J_{c-\frac{1}{2}}(\frac{m^{(n)}}{k})}/
{Y_{c-\frac{1}{2}}(\frac{m^{(n)}}{k})} $.
Due to the Bessel function relation of
\begin{eqnarray}
\label{eq:besseld}
    \left( \partial_z +  \frac{c}{z}\right)
f_L^{(n)}(z) &=& \frac{m^{(n)}\sqrt{z}}{N_L^{(n)}}
\left[J_{c-\frac{1}{2}}(m^{(n)} z) +
\beta_L^{(n)} Y_{c-\frac{1}{2}}(m^{(n)} z)\right], \\
\nonumber
 \left( \partial_z - \frac{c}{z}\right)f_R^{(n)}(z)
 &=& -\frac{m^{(n)}\sqrt{z}}{N_R^{(n)}}
\left[J_{c+\frac{1}{2}}(m^{(n)} z) + \beta_R^{(n)}
Y_{c+\frac{1}{2}}(m^{(n)} z)\right]
\,,
\end{eqnarray}
Eq.~(\ref{eq:doddzero}) yields
\begin{equation}
\beta_L^{(n)}= -\frac{J_{c-\frac{1}{2}}(\frac{m^{(n)}}{k})}
{Y_{c-\frac{1}{2}}(\frac{m^{(n)}}{k})} =\beta_R^{(n)} .
\label{even}
\end{equation}
It is clear that $N_L^{(n)}=-N_R^{(n)}$ from Eq.~(\ref{eq:besseld}).

Without the localized fermion mass,
KK masses of a bulk fermion
depend on its $\Z \times \Z'$ parity
and the bulk Dirac mass parameter $c$.
Under the $\Z \times \Z'$ symmetry,
a generic 5D bulk fermion can have the following four different
transformation property:
\begin{eqnarray}
\label{eq:Psi1}
\hpsi_1(x,y) & = &
\sqrt{k}\sum_n [\psn_{1L}(x) \fn_{1L(++)}(y)
+ \psn_{1R}(x) \fn_{1R(--)}(y) ] \,,\\
\label{eq:Psi2}
\hpsi_2(x,y) & = &
\sqrt{k}\sum_n [\psn_{2L}(x) \fn_{2L(--)}(y)
+ \psn_{2R}(x) \fn_{2R(++)}(y) ] \,,\\
\label{eq:Psi3}
\hpsi_3(x,y) & = &
\sqrt{k}\sum_n [\psn_{3L}(x) \fn_{3L(+-)}(y)
+ \psn_{3R}(x) \fn_{3R(-+)}(y) ] \,,\\
\label{eq:Psi4}
\hpsi_4(x,y) & = &
\sqrt{k}\sum_n [\psn_{4L}(x) \fn_{4L(-+)}(y)
+ \psn_{4R}(x) \fn_{4R(+-)}(y) ]\,,
\end{eqnarray}
where
the $\Z\times \Z'$ parities
are denoted by $(PP')$ in $\fn_{(PP')}$.
Note that
the $P\,(P')$ of $f_{iL}^\n$
is opposite to that of $f_{iR}^\n$.

With Eq.~(\ref{eq:betaN}), the mode functions are
\bea
f_{iL}^\n(z) &=&
\frac{\sqrt{z}}{N_i^\n}
\left[
J_{c_i+1/2}(m_i^\n z)+\beta_i^\n Y_{c_i+1/2}(m_i^\n z)
\right],\nonumber \\
f_{iR}^\n(z) &=&-
\frac{\sqrt{z}}{N_i^\n}
\left[
J_{c_i-1/2}(m_i^\n z)+\beta_i^\n Y_{c_i-1/2}(m_i^\n z)
\right]
\,.
\label{mode1}
\eea
The coefficient $\beta_i^\n$ is fixed
by the fact that a $\Z\,(\Z')$-odd function vanishes at the corresponding boundary.
For example, the $f_{1R(--)}^\n$ and $f_{2L(--)}$
vanish at both UV and IR branes,
which doubly constrain the $\beta_i^\n$:
\bea
\label{eq:beta1}
\beta_1^\n &= & - \dfrac{J_{c_1-1/2}(m^\n/k)}{Y_{c_1-1/2}(m^\n/k)}=
     - \dfrac{J_{c_1-1/2}(m^\n/T)}{Y_{c_1-1/2}(m^\n/T)}\,,
      \\ \label{eq:beta2}
     \beta_2^\n &= & - \dfrac{J_{c_2+1/2}(m^\n/k)}{Y_{c_2+1/2}(m^\n/k)}=
     - \dfrac{J_{c_2+1/2}(m^\n/T)}{Y_{c_2+1/2}(m^\n/T)}\,.
\eea
Similarly we have
\bea
\label{eq:beta3}
    \beta_3^\n &= & - \dfrac{J_{c_3-1/2}(m^\n/k)}{Y_{c_3-1/2}(m^\n/k)}=
     - \dfrac{J_{c_3+1/2}(m^\n/T)}{Y_{c_3+1/2}(m^\n/T)}\,,
  \\ \label{eq:beta4}
     \beta_4^\n &= & - \dfrac{J_{c_4+1/2}(m^\n/k)}{Y_{c_4+1/2}(m^\n/k)}=
     - \dfrac{J_{c_4-1/2}(m^\n/T)}{Y_{c_4-1/2}(m^\n/T)}
\,.
\eea

\begin{figure}[tbh]
  \includegraphics[scale=1]{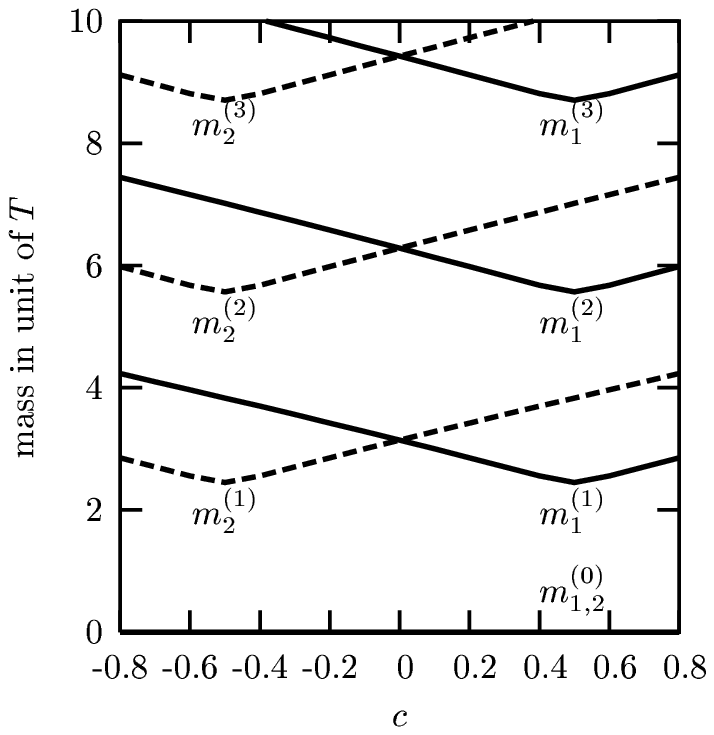}
  \includegraphics[scale=1]{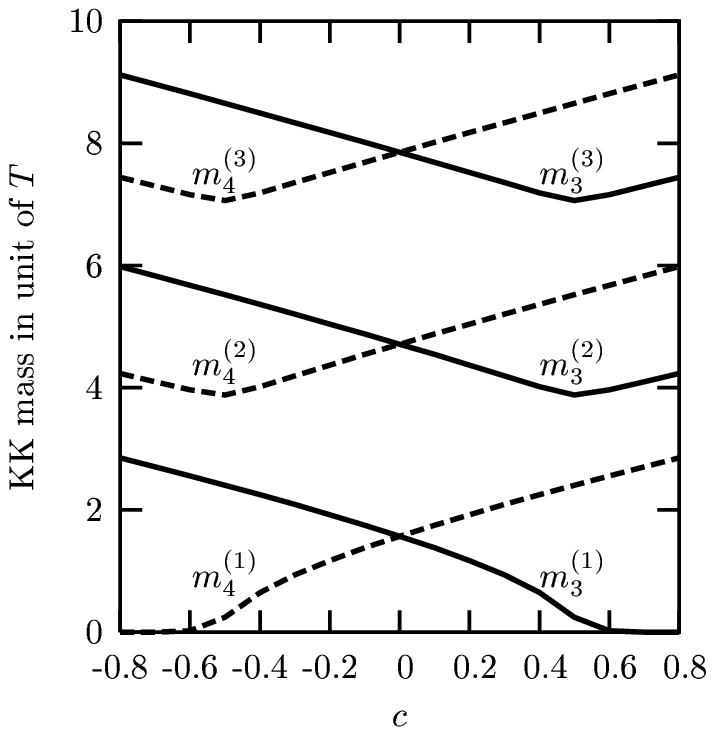}
  \caption{\label{fig:bulkf0}
KK mass spectra of a bulk fermion $\Psi_i$ in unit of $T$
as a function of the bulk mass parameter $c$.
The $\Z\times \Z'$ parities of each fermion is
described in the text.}
\end{figure}

In Fig.~\ref{fig:bulkf0},
we present the KK masses of a bulk fermion $\Psi_i$
in unit of $T$
as a function of the bulk mass parameter $c$.
It is clear that
$\Psi_{1}(\supset\Psi_{1L}^{(++)})$ and
$\Psi_{2}(\supset\Psi_{2R}^{(++)})$
can accommodate zero modes.
An unexpected feature is that the zero mode mass of
$\Psi_{3}$ ($\Psi_{4}$) can be considerably light
for $c_3>0.5~(c_4<-0.5)$.

\section{Bulk fermion with Yukawa interaction on the brane}
\label{sec:bfermionYukawa}

The accommodation of the SM fermions in the bulk RS theories
has some delicate features.
First a single SM fermion with left- and right-handed chirality should be described
by \emph{two} 5D Dirac fermions.
For example, the left-handed up quark is to be described by the $\Psi_1$ type
while the right-handed up quark by the
$\Psi_2$ type.
Another interesting problem is
the generation of light and realistic masses
for the SM fermions~\cite{Gherghetta:2000qt}.
Even though
the KK zero modes
are good candidates for the SM fermions,
their zero masses should be lifted only a little.
In the ADMS model,
the localized Higgs VEV plays this role without explicit breaking of
gauge symmetry~\cite{Agashe:2003zs}.
Different strengths of Yukawa couplings can explain diverse mass spectrum
of the SM fermions as in the SM.
In the Higgsless model,
it is also possible to get realistic SM fermion masses
by boundary conditions~\cite{Csaki:2003sh}.
Unfortunately the basic set-up is somewhat complicated for each SM fermion:
A 4D gauge invariant Dirac mass is to be introduced,
which is localized on the IR brane and mixes the $\Psi_1$ and $\Psi_2$ types;
a new 4D Dirac spinor, localized on the UV brane, is also required to
mix with the $\Psi_2$ type fermion.

For simplicity, we consider the case
where SM fermion masses are generated by
the localized Yukawa couplings
between two fermion fields of $\Psi_1$
and $\Psi_2$.
The five-dimensional fermion action is
\begin{eqnarray}
S_{\rm fermion}=\int d^4x \int^{L}_0 dy\left[
\overline{\hPsi}_i e^\sigma i\gamma^\mu\partial_\mu
\hPsi_i -\frac{1}{2} \overline{\hPsi}_i \gamma_5\partial_y
\hPsi_i +\frac{1}{2}(\partial_y
\overline{\hPsi}_i)\gamma_5 \hPsi_i
 + m_{iD} \overline{\hPsi}_i\hPsi_i \right],
\end{eqnarray}
where $\sum_{i=1}^2$ is assumed for repeated index $i$.
The Yukawa interaction localized
on the IR brane couples
the $\Psi_1$ with $\Psi_2$:
\begin{eqnarray}
\label{eq:Syukawa}
S_{\rm Yukawa}=-\int d^4 x \, dy  \lm
\left(\overline{\hpsi}_{1}
\hpsi_{2}+\overline{\hpsi}_{2}
\hpsi_{1}\right)\delta(y-L) .
\end{eqnarray}
Here, $\lambda_v\equiv\lambda_{5}\langle H \rangle/T$,
$\lambda_{5}$ is the 5D
dimensionless Yukawa coupling  and $H(x)$ is
a canonically normalized Higgs
scalar defined by $H(x)=\es H_5(x)$.
The 5D total action becomes $
S_{5D}=S_{\rm fermion}+S_{\rm Yukawa}
$.

To simplify $S_{5D}$, a technical  problem arises as
the Dirac delta function is positioned at $y=L$
while the $y$-integration range is
in $[0, L]$~\cite{Csaki:2003sh}.
We regulate this by using the periodicity of $S^1$ space
and dividing the integration into
\begin{equation}
\label{eq:Sdiv}
S_{5D} =
\int^{L-\ves}_{0} d y\  S_{4D} +
 \frac{1}{2} \int^{L+\ves}_{L-\ves} d y\  S_{4D}
 \equiv S_{\rm bulk} + S_\partial .
\end{equation}
For a small
positive $\varepsilon$,
$S_{\rm bulk}$
is well-defined, given by
\begin{eqnarray}
\label{eq:Sbulk0}
S_{\rm bulk} &=&
\int d^4x\left\{ \int^{L-\ves}_{0} dy
\,\left[
\overline{\hPsi}_i \left( e^\sigma i\gamma^\mu\partial_\mu  -
\gamma_5\partial_y
 + m_{iD} \right)\widehat\Psi_i \right]
 +
 \left.\frac{1}{2}\overline{\hPsi}_i \gamma_5 \hPsi_i
 \right|^{L-\ves}_0 \right\}\,,
\end{eqnarray}
where $f(x)|_a^b \equiv f(b)-f(a)$ and $S_\partial $ is
\beq
\label{eq:Srd0}
S_\partial =
\int d^4x\left\{ -\frac{1}{2}
\int^{L+\ves}_{L-\ves} dy \,\overline{\hPsi}_i
\gamma_5\partial_y \hPsi_i
 -\frac{1}{2}\,\lambda_v \left.\left(\overline{\hPsi}_1 \hPsi_2 +
\overline{\hPsi}_2 \hPsi_1 \right)\right|_{y=L}
+
\left.\frac{1}{4}\,\overline{\hPsi}_i \gamma_5 \hPsi_i
 \right|^{L+\ves}_{L-\ves}
 \right\}\,.
\eeq

The absence of the localized fermion mass on the Planck brane guarantees
the continuity of
the mode functions at $y=0$:
\beq
\label{eq:continuity}
\left.\overline{\hPsi}_i \gamma_5 \hPsi_i \right|_{y=0} =0
\,.
\eeq
Furthermore the $\Z'$-oddity of
$\overline{\hPsi}_i \gamma_5 \hPsi_i$ implies
\begin{equation}
\label{eq:odd-psi-gm5}
\left.\overline{\hPsi}_i \gamma_5 \hPsi_i
\right|^{L+\ves}_{L-\ves} = \left.-2\,\overline{\hPsi}_i
\gamma_5 \hPsi_i \right|_{y=L-\ves}
\,.
\end{equation}
Equations (\ref{eq:continuity}) and (\ref{eq:odd-psi-gm5})
give rise to the cancelation
between the last terms
of Eqs.~(\ref{eq:Sbulk0})
and (\ref{eq:Srd0}).
Therefore, the $S_{\rm bulk}$ is
\begin{eqnarray}
S_{\rm bulk}
 &=& \int\! d^4x\!
\int^{L-\ves}_{0}\hspace{-2mm} dy \left[e^\sigma(
\overline{\hPsi}_{iL} i\gamma^\mu\partial_\mu \hPsi_{iL}+
\overline{\hPsi}_{iR} i\gamma^\mu\partial_\mu \hPsi_{iR})
\right.
\\ \nonumber
&&
\left.-\overline{\hPsi}_{iL}(\partial_y-m_{iD})\hPsi_{iR}
  +\overline{\hPsi}_{iR}(\partial_y+m_{iD})\hPsi_{iL}
  \right]
  \,.
\eea

More comments on simplifying $S_{\partial}$
are in order here.
The $\Z'$-even parity of $\hPsi_{1L}$ and
$\hPsi_{2R}$ guarantees the continuity at $y=L$,
which eliminates the infinitesimal integration of
$\overline{\hPsi}_{1R}\gm_5 \partial_y \hPsi_{1L}$
and $\overline{\hPsi}_{2L}\gm_5 \partial_y \hPsi_{2R}$
in Eq.~(\ref{eq:Srd0}).
On the contrary, the presence of Yukawa term hints the
discontinuity of $\Z'$-odd $\hPsi_{1R}$ and
$\hPsi_{2L}$ at $y=L$.
Nevertheless at $y=L$ the values of $\Z'$-odd functions can be assigned
zero,
which is possible by setting zero
the $y=L$ boundary value of periodic $\mathrm{sign}(y)$ function
in the bulk Dirac mass term~\footnote{This can be easily seen
by the equation of motion in terms of $y$ coordinate,
give by $\partial_y f_{1L}+m_{1D}f_{1L} = -m^\n e^\sigma f_{1R}$}.
Among Yukawa terms in Eq.~(\ref{eq:Srd0}),
therefore, $(\overline{\hPsi}_{1R}\hPsi_{2L}+h.c.)|_{y=L}$
vanishes.
Finally integration by part and Eq.~(\ref{eq:odd-psi-gm5})
simplifies $S_{\partial}$ as
\bea
S_{\partial}&=&
\int
d^4x
\left[\,\left.
\left(
\overline{\hPsi}_{1L}\hPsi_{1R}- \overline{\hPsi}_{2L}\hPsi_{2R}
\right)\right|_{y=L-\ves}
-\frac{\lambda_v}{2}
\left.\left(\overline{\hPsi}_{1L} \hPsi_{2R} + \overline{\hPsi}_{2R}
\hPsi_{1L} \right)\right|_{y=L}
\right]
\, .
\eea
The variation of $S_{\rm bulk}$ gives equations of
motion for bulk fermions, while $
S_{\partial } =0$ gives boundary conditions.

Without Yukawa terms,
$\Psi_1$ and
$\Psi_2$ have their own KK mass spectra,
determined by the bulk Dirac mass parameter $c_i$.
As the Yukawa couplings turn on between $\Psi_1$ and $\Psi_2$,
$\psi_{1L}^{(n)}$ and $\psi_{1R}^{(n)}$
mix with $\psi_{2L}^{(n)}$ and $\psi_{2R}^{(n)}$, respectively.
Denoting the KK mass eigenstates by
$\chi^{\n}_{L,R}(y)$,
the KK expansion of the bulk field is
\begin{equation}
\label{eq:KKexpansion:mass}
\hat{\Psi}_{i} =
\sqrt{k}\sum_n \left[ \chi_{L}^{(n)}(x) f_{iL}^{(n)}(y)
+ \chi_{R}^{(n)}(x) f_{iR}^{(n)}(y)
\right]
.
\end{equation}
With the modified normalization of
\begin{equation}
\sum_{i=1,2}k\int_0^L d y\,
e^\sigma  { f}^{(n)}_{iL}  { f}^{(m)}_{iL}
=
k\sum_{i=1,2}\int_0^L d y\,
e^\sigma  { f}^{(n)}_{iR}  { f}^{(m)}_{iR}
= \delta_{nm}
\,,
\end{equation}
and the equations of motion of
\begin{eqnarray}
\label{eq:eomYukawa}
 \partial_y  f_{iR}^{(n)} - m_{iD}  f_{iR}^{(n)}
 & = & m^{(n)} e^\sigma  f_{iL}^{(n)}\,,  \nonumber\\
 -\partial_y  f_{iL}^{(n)} - m_{iD}  f_{iL}^{(n)}
  & = & m^{(n)} e^\sigma  f_{iR}^{(n)} ,
\end{eqnarray}
the 4D
effective action consists of the KK fermions:
\begin{equation}
\label{eq:Seff-chi}
S_{\rm eff}
 = \int d^4 x \sum_n\left[\overline{\chi}^{(n)}_L i \gamma^\mu \partial_\mu
 \chi^{(n)}_L + \overline{\chi}^{(n)}_R i\gamma^{\mu} \partial_\mu
\chi^{(n)}_R - m^{(n)}(\overline{\chi}^{(n)}_L \chi^{(n)}_R +
\overline{\chi}^{(n)}_R \chi^{(n)}_L)
     \right] \ .
\end{equation}
The general solutions of Eq.~(\ref{eq:eomYukawa}) are the same as Eq.~(\ref{mode1}).

The substitution of Eq.~(\ref{eq:KKexpansion:mass})
into $S_\partial=0$
with the continuity of $\Z'$-even functions at $y=L$, we have
\beq
 f_{1R}^{(n)}|_{y=L-\ves} =
 \frac{\lambda_v}{2}  f_{2R}^{(n)}|_{y=L}
 \,,\quad
 f_{2L}^{(n)}|_{y=L-\ves}= -
 \frac{\lambda_v}{2}  f_{1L}^{(n)}|_{y=L}. \label{bcf}
\eeq
In the followings, we will ignore infinitesimal $\varepsilon$ and
consider only $z$ coordinates.
Finally we have all boundary conditions
at
$\zu=1/k$ and $\zi=1/T$:
\begin{eqnarray}
\label{eq:BCf1R}
\left. f_{1R}^{(n)} \right|_\frac{1}{k}  =  0 ,&& \left.
f_{1R}^{(n)} \right|_\frac{1}{T}  =
\left.  \frac{\lambda_v}{2}  f_{2R}^{(n)} \right|_\frac{1}{T},   \\
\label{eq:BCf2L}
\left. f_{2L}^{(n)} \right|_\frac{1}{k}  =  0, && \left.
f_{2L}^{(n)} \right|_\frac{1}{T}  =
-\left.  \frac{\lambda_v}{2}  f_{1L}^{(n)} \right|_\frac{1}{T} , \\
\label{eq:BCf1L}
\left.(\partial_z + \frac{c_1}{z}) f_{1L}^{(n)}
\right|_\frac{1}{k}  =  0 ,&& \left.(\partial_z + \frac{c_1}{z})
f_{1L}^{(n)} \right|_\frac{1}{T}  =
\left.  \frac{\lambda_v}{2} m^{(n)} f_{2R}^{(n)} \right|_\frac{1}{T},   \\
\label{eq:BCf2R}
\left.(\partial_z - \frac{c_2}{z}) f_{2R}^{(n)}
\right|_\frac{1}{k}  =  0, && \left.(\partial_z - \frac{c_2}{z})
f_{2R}^{(n)} \right|_\frac{1}{T}  = \left. \frac{\lambda_v}{2}
m^{(n)} f_{1L}^{(n)} \right|_\frac{1}{T}  .
\end{eqnarray}

At $z=1/k$ the relations are the same as the case
of $\lm=0$, yielding
\bea
\label{eq:betaLR}
\beta_1^\n\equiv
\beta_{1R}^{(n)}&=&\beta_{1L}^{(n)}=
-\frac{J_{c_1-\frac{1}{2}}(\frac{m^{(n)}}{k})}
{Y_{c_1-\frac{1}{2}}(\frac{m^{(n)}}{k})}
, \\
\nonumber
\beta_2^\n \equiv
\beta_{2L}^{(n)}&=&\beta_{2R}^{(n)}=
-\frac{J_{c_2+\frac{1}{2}}(\frac{m^{(n)}}{k})}
{Y_{c_2+\frac{1}{2}}(\frac{m^{(n)}}{k})} .
\eea
As in the previous section,
the normalization factors of the left-handed and right-handed mode functions
are related by
\bea
N_1^{(n)}\equiv
N_{1L}^{(n)}=-N_{1R}^{(n)},
\quad
N_2^{(n)}\equiv
-N_{2L}^{(n)}=N_{2R}^{(n)}
\,.
\eea

The boundary conditions at $z=1/T$ in Eqs.~(\ref{eq:BCf1R})
and (\ref{eq:BCf2L}) give
\begin{eqnarray}
\label{eq:BCy1}
\frac{1}{N_1^{(n)}} \left[
J_{c_1-\frac{1}{2}}(x^\n) +\beta_1^{(n)}
Y_{c_1-\frac{1}{2}}(x^\n)\right] &=  &-
\frac{\lambda_v}{2N_2^{(n)}} \left[
J_{c_2-\frac{1}{2}}(x^\n) +\beta_2^{(n)}
Y_{c_2-\frac{1}{2}}(x^\n)\right]\,,  \\
\label{eq:BCy2}
\frac{1}{N_2^{(n)}} \left[
J_{c_2+\frac{1}{2}}(x^\n) +\beta_2^{(n)}
Y_{c_2+\frac{1}{2}}(x^\n)\right] &=  &\phantom{-}
\frac{\lambda_v}{2N_1^{(n)}} \left[
J_{c_1+\frac{1}{2}}(x^\n) +\beta_1^{(n)}
Y_{c_1+\frac{1}{2}}(x^\n)\right] \, ,
\label{eq2}
\end{eqnarray}
where $x^\n=m^\n/T$.
The elimination of $N_1^{(n)}$ and $N_2^{(n)}$ by multiplying Eq.
(\ref{eq:BCy1}) and Eq.~(\ref{eq:BCy2})
and the substitution of $\beta_{1,2}^\n$ in Eq.~(\ref{eq:betaLR})
produce the final master equation:
\begin{equation}
{\cal J}^{--}_1(x^{(n)}) {\cal J}^{++}_2(x^{(n)})
=-\frac{\lambda_v^2}{4} {\cal J}^{-+}_1(x^{(n)}) {\cal
J}^{+-}_2(x^{(n)}), \label{mass1}
\end{equation}
where ${\cal J}^{\pm\pm'}_i$ is defined by
\begin{eqnarray}
{\cal J}^{\pm\pm'}_i(x^{(n)}) &=&
Y_{c_i\pm\frac{1}{2}}\left(\es \,x^\n \right)
J_{c_i\pm'\frac{1}{2}}\left( x^\n \right)
-
J_{c_i\pm\frac{1}{2}}\left(\es \,x^\n \right)
Y_{c_i\pm'\frac{1}{2}}\left(x^\n\right)
\,.
\label{pm}
\end{eqnarray}

These master equations clearly show
the relation of the $\ZZ$ parity
and the large localized Higgs VEV as in the gauge boson case.
When $\lambda_v$ is zero,
the KK spectra of $\hpsi_1$ and $\hpsi_2$ are
the same  as in the previous section.
As $\lambda_v\rightarrow \infty$
the right-handed side of Eq.~(\ref{eq:BCy2}) should vanish,
yielding
\beq
\beta_1^\n |_{\lm \to \infty} =
-\frac{J_{c_1+1/2}(m^\n/T)}{Y_{c_1+1/2}(m^\n/T)}
=-\frac{J_{c_1-1/2}(m^\n/k)}{Y_{c_1-1/2}(m^\n/k)}
\,,
\eeq
where the second equality comes from Eq.~(\ref{eq:betaLR}).
This is identical to the $\beta^\n_3$ in Eq.~(\ref{eq:beta3}) except for $c_i$.
The KK mass spectrum of $\Psi_1$ in the large $\lm$ limit
is the same as that of $\Psi_3$ without $\lm$: $\Psi_1$ mimics $\Psi_3$.
Similarly, the
$\Psi_2$ mimics $\Psi_4$.

\begin{figure}[tb]
  \includegraphics[scale=1]{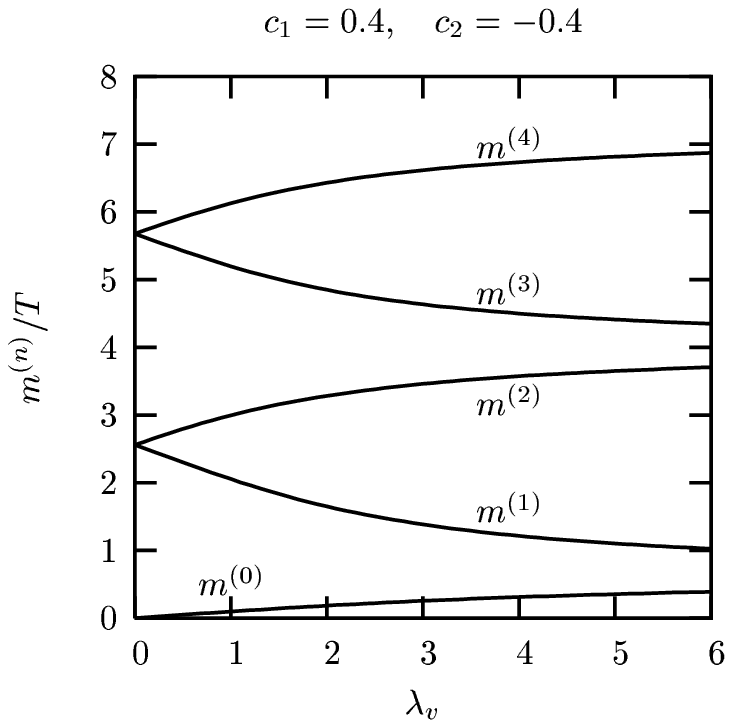}
  \includegraphics[scale=1]{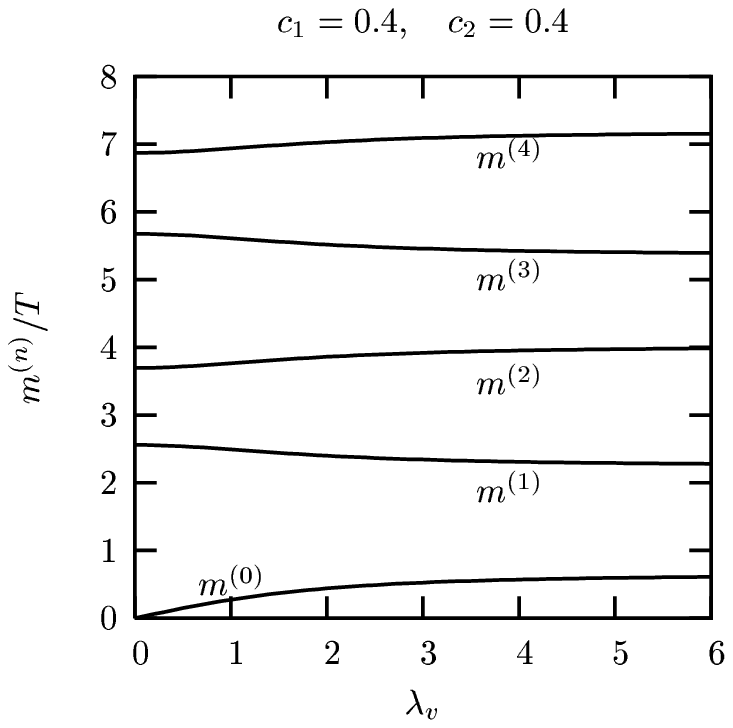}
  \caption{\label{fig:bulkfY}
KK mass spectra of bulk fermions
$\Psi_1(\supset \Psi_{1L}^{(++)})$ and
$\Psi_2(\supset \Psi_{2R}^{(++)})$ in unit of $T$
as a function of Yukawa mass term $\lm$.
}
\end{figure}

Figure \ref{fig:bulkfY} shows the KK masses of $\Psi_1$ and $\Psi_2$
as a function of $\lm$.
We present the numerical results for two cases,
$[c_1=0.4,c_2=-0.4]$
case and
$[c_1=0.4,c_2=0.4]$ case.
As the Yukawa term increases,
the zero mode acquires non-zero mass $m^{(0)}$.
For large $\lm$,
the $m^{(0)}$ becomes saturated as in the gauge boson case,
since the KK mode become a mixed state of higher KK modes.
Another interesting feature is
that the first KK mode mass \emph{decreases} as $\lm$ increases,
contrary to the bulk gauge boson case where
$m_A^{(1)}$ increases with $\ai$
(see Fig.~\ref{fig:bulkAair}).
The mass drop due to $\lm$ is maximal when
the two KK mass spectra were degenerate at $\lm=0$, e.g.,
$[c_1=0.4,c_2=-0.4]$ case.
This KK mode degenerate case will
leave dramatic signatures at high energy colliders:
The KK modes of light quarks show doubly degenerate mass spectrum while
the first KK mode of top quark can be considerably light.
It is very feasible, therefore,
that the first signal of KK fermions comes from
the top quark mode which alone possesses non-negligible
Yukawa mass.
The saturation of zero mode mass and the dropping of first excited KK mode mass
are consistent with the existence of two light KK mode in
the transition from $\hpsi_1$ to
$\hpsi_3$ and $\hpsi_2$ to $\hpsi_4$ spectra
for the $\lambda_v\rightarrow \infty$ limit.

\section{Gauge coupling universality}
\label{sec:universality}
In the previous sections, it is shown that
the presence of the localized mass terms
generates non-zero masses for the zero modes
as well as modifying other higher KK mode masses.
Another important influence of localized mass terms
is on mode functions.
Without localized mass terms,
the zero mode functions of a bulk gauge boson
($\tilde f_A^{(0)}$) and
a $\Psi_1$-type fermion ($\tilde f_{L,R}^{(0)}$)
are
\bea
\tilde{f}_{A}^{(0)} &=& \frac{1}{\sqrt{k L}},
\\ \no
\tilde{f}_{L(++)}^{(0)} &=&  \frac{(k z)^{-c}}{N^{(0)}}
\,,
\quad
\tilde{f}_{R(--)}^{(0)}  = 0,
\eea
where the tildes over mode functions emphasize the absence of boundary mass terms.
The localized masses
change these functional forms.
Since our four-dimensional effective gauge coupling
is obtained by convoluting mode functions of a gauge boson
and two fermions,
different changes of mode functions by different localized masses can deviate
the SM relations of gauge couplings.
In what follows, we focus on a simple scenario
where only the top quark
Yukawa coupling is non-zero.
If the 4D gauge coupling $g$ is defined by
the $u$-$d$-$W$ coupling,
the top-bottom-$W$ coupling,
$g_{Wtb}$, departs from $g$ due to the deformed mode functions:
The gauge coupling universality may be in danger.

In the five dimensional RS theory,
the changed current interaction of $SU(2)_L$
is
\beq
S_{CC} = \int d^4x \int d z
\frac{i g_5}{\sqrt{2 k}}\,
\overline{\hat{q}}_u(x,z) \Wsh^+ (x,z) \hat{q}_d(x,z) + H.c.,
\eeq
where $g_5$ is the dimensionless 5D gauge coupling,
$Q = (q_u, q_d)^T$ is a $SU(2)_L$ doublet.
The $q_u(x,y)$ and $q_d(x,y)$ are $\Psi_1$ type in Eq.~(\ref{eq:Psi1}),
i.e.,
$q_{uL}$ and $q_{dL}$ have $(++)$
parity.
The five dimensional gauge coupling $g_5$ is related with
four dimensional gauge coupling $g$ by
\beq
\label{eq:gg5NO}
g \equiv g_5 \int k d z f_{q_u L}^{(0)} f_{q_d L }^{(0)} {f}_W^{(0)}
\,.
\eeq
If the localized gauge boson mass is absent so that $\tilde{f}_{W}^{(0)}$ is constant,
the fermion mode function normalization in Eq.~(\ref{eq:normf})
guarantees
the same gauge coupling strength,
irrespective of the localized Yukawa coupling strength.
Gauge coupling universality remains intact.

As the localized gauge mass terms turn on,
non-constant $f_W^{(0)}$
leads to different relations between $g$ and $g_5$ according to $\lm$.
We define the four-dimensional gauge coupling $g$
by the $W$-$u$-$d$ coupling
with the up and down quark Yukawa couplings
neglected:
\beq
\label{eq:g-def}
g \equiv g_5 \int k d z
\tilde{f}_{q_u L}^{(0)} \tilde{f}_{q_d L }^{(0)} f_W^{(0)}
\,.
\eeq
Substantial top quark Yukawa coupling $\lambda_t$
changes the mode function $f_{tL}^{(0)}$
and thus the $W$-$t$-$b$ gauge coupling $\gtb$.
Note that the assumption of $\lambda_b=0$
leads to $\tilde{f}_{bR}^{(0)}=0$,
eliminating anomalous right-handed $Wtb$ coupling.
The degree of gauge coupling universality violation is defined by
\beq
\delta g_{Wtb} =\frac{g_{Wtb}}{g}-1
=
\frac{\int k d z
{f}_{q_t L}^{(0)} \tilde{f}_{q_b L }^{(0)} f_W^{(0)}}{\int k d z
\tilde{f}_{q_u L}^{(0)} \tilde{f}_{q_d L }^{(0)} f_W^{(0)}}-1
\,.
\eeq

For the numerical evaluation of $\delta g_{Wtb}$,
let us discuss the model parameters.
First we have the effective electroweak scale $T$.
Since the up and down quarks in a given $SU(2)_L$ doublet
should shared the same
bulk Dirac mass, we have two bulk Dirac mass parameter
for the first and third generation,
denoted by $c$ and $c_t$.
Non-zero $\ai$ and $\lambda_t$
are traded with
the observed $m_W$ and $m_{\rm top}$.
In summary, the following three parameters determine
$\delta g_{Wtb}$:
\beq
T, \quad  c,\quad  c_t
\,.
\eeq

Figure \ref{fig:gwtb} shows the $\delta g_{Wtb}$ as a function of $T$.
It can be easily seen that
the deviation decreases with increasing $t$,
and is negligible unless
$c$ is not too different from $c_t$.
In particular, the $c=c_t$ case practically preserves
the gauge coupling universality.
However, if $|c-c_t|$ becomes substantial (e.g., $[c=-c_t=0.4]$ case),
the deviation can be a few percent for $T \simeq 1.5 \mathrm{TeV}$.
Concerning the KK mass spectra,
the $c \approx -c_t$ case allows
substantially light KK mass of the first KK mode
as $\lm$ increases.
On the contrary,
the $c\approx c_t$ case implies almost negligible $\delta g_{Wtb}$
even for relative light $T$
whose KK mass spectra are quite different
for $\Psi_1$ and $\Psi_2$.
In conclusions,
the parameter space which guarantees gauge coupling universality
has the KK mass spectrum which
is similar
with the KK masses without Yukawa terms.

\begin{figure}
  \includegraphics[scale=1]{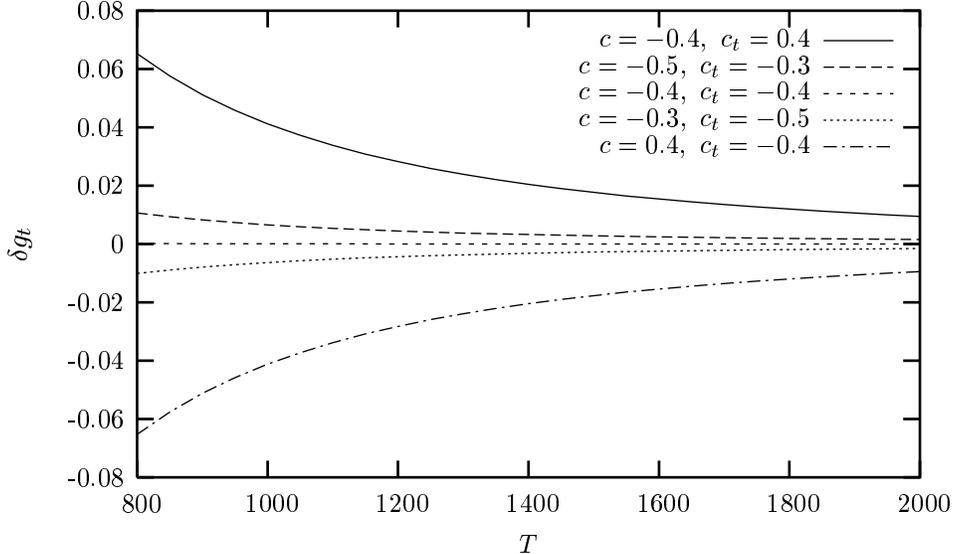}
  \caption{\label{fig:gwtb}
The degree of the gauge coupling non-universality,
defined by $\delta g_{Wtb} =g_{Wtb}/g-1$,
as a function of $T$ for various combinations
of $c$ and $c_t$.
}
\end{figure}

Even though the violation of gauge coupling unification is, if any,
a breakthrough in particle physics,
its magnitude with $T$ around a few TeV is below
a few percent.
At a hadron collider like LHC, it is too small to detect.
If its correlation with other physical observable
such as KK masses is strong enough,
it can be a valuable information.
Restricting ourselves to the KK mode degenerate case (i.e., $c_{qL}=-c_{qR}$),
we plot
the correlation between $\delta g_{Wtb}$ and
the mass difference of the first KK modes of light quark and top quark
in unit of their mass sum in Fig.~\ref{fig:cor}.
The effective electroweak scale $T$ is fixed to be 2 TeV,
while the parameter space of $c$ and $c_T$ in $[-0.4,\,0.4]$
are all scanned.
The parameters $\lm$ and $\ai$  are determined by the SM top quark
and $W$ boson mass, respectively.
As can be seen in Fig.~\ref{fig:cor},
we do have quite significant correlation.
In the parameter space where the first KK mode
of top quark is lighter than that of light quarks,
$\delta g_{Wtb}$ is negative.
\begin{figure}
  \includegraphics[scale=1]{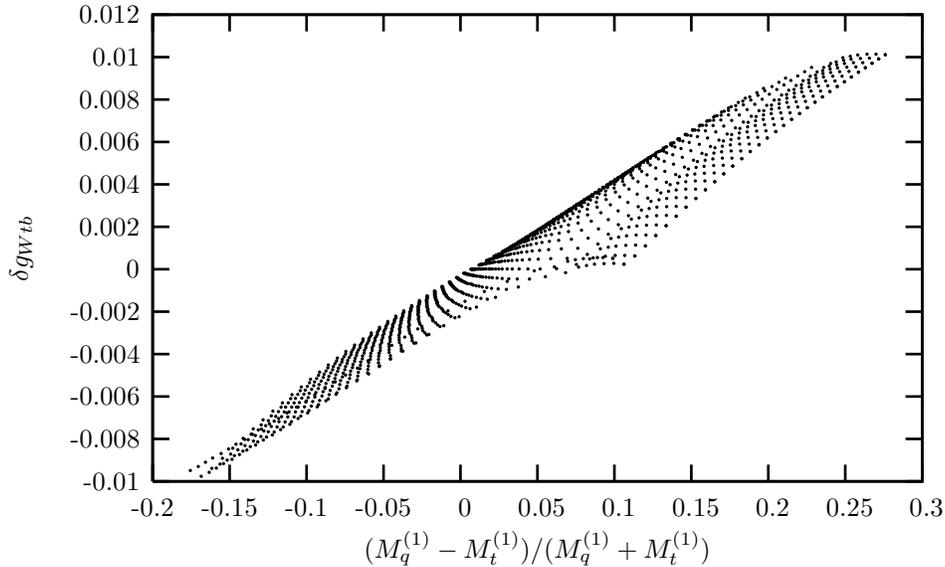}
  \caption{\label{fig:cor}
In the KK mode degenerate case,
we plot the correlation between the $\delta g_{Wtb} =g_{Wtb}/g-1$
and the mass difference of the first KK modes of light quark and top quark.
The effective electroweak scale $T$ is fixed to be 2 TeV,
and $c,~c_T \in [-0.4,\,0.4]$.
}
\end{figure}

\section{Conclusion}
\label{sec:conclusion}
We have studied master equations for the Kaluza-Klein (KK)
masses of a bulk gauge boson and a bulk fermion
in a five-dimensional (5D) warped space compactified on a
$S^1/\Z\times\Z'$ orbifold.
These master equations accommodate the general case with
the brane-localized and bulk mass terms.
Comprehensive understanding for
the KK mass spectra and their behavior
is crucial to verify and discriminate the Higgsless model from
the ADMS model.

After presenting master equations for the bulk gauge boson,
it is explicitly shown that
the Neumann boundary
condition (for $\Z$-even parity) in the limit of large localized mass
is equivalent to the Dirichlet
boundary condition (for $\Z$-odd parity).
This correspondence relates among KK mass spectra
of gauge bosons with different $\ZZ$ parities.
A bulk gauge boson with $(++)$ parity and very large localized mass
on the UV brane (denoted by $\au$)
has the same KK mass spectrum
with a $(-+)$ bulk gauge boson without any localized mass.
In brief, the $(++)$ gauge field in the large $\au$ limit
mimics $(-+)$ gauge field.
Similarly, a $(++)$ bulk gauge boson in the large $\ai$ limit
mimics a $(+-)$ bulk gauge boson without any localized mass.
This implies that
the ADMS model
in the large limit of the Higgs VEV
can be related to the Higgsless model.
Thus we can understand why one cannot avoid TeV-scaled KK
states even in the case of infinite VEV of the localized Higgs
boson(s).
This is a generic property of the gauge theory in the
truncated AdS space with TeV-valued boundary.
Through numerical calculations,
we have presented the KK masses of a bulk gauge boson
for various $\ZZ$ parities.
The first KK mode with $(+-)$ parity
is shown to be remarkably light with mass of order 100 GeV.
The $\ai$-dependence of the KK masses for $(++)$ parity is also
presented:
With increasing
$\ai$,
not only does the zero mode KK mass acquire non-zero mass,
but the first KK mass also increase.
We have also shown that the method of $\au$ to raise the zero mode mass is different:
As soon as the $\au$ above $\sim 10^{-15}$,
the zero mode mass jumps to the TeV scale.

We have extended this discussion to the bulk fermion case.
A bulk fermion on a $\SZZ$ orbifold has four different boundary
parities: The parity of the left-handed chiral fermion can be
$(++)$, $(+-)$, $(-+)$, and $(--)$,
while the parity of the corresponding right-handed fermion is opposite.
First we have considered a simple case without any localized mass term.
Through numerical calculations, it was shown that
the first KK mode mass of $\Psi_3 \supset \Psi_{3L}^{(+-)}$
and $\Psi_4 \supset \Psi_{4L}^{(-+)}$
is substantially light
for $c_3>0.5$ and $c_4<-0.5$, respectively.
In order to explain the SM fermion masses,
we have introduced
the brane localized Yukawa coupling between
$\Psi_1 \supset \Psi_{1L}^{(++)}$ and
$\Psi_2 \supset \Psi_{2R}^{(++)}$.
From the coupled boundary conditions,
the final master equations are derived for the KK masses of
bulk fermions.
Similar correspondence between the $\ZZ$ parity and the large
localized Yukawa coupling is also shown:
The $\Psi_1$ ($\Psi_2$) as $\lm\to\infty$ mimics the
$\Psi_3 $ ($\Psi_4 $)
with $\lm=0$.
Another interesting feature is that the first KK mode
mass decreases with increasing Yukawa coupling.
In the future collider, the top quark KK mode
is one of the first candidates to be detected.

Finally we have investigated the violation of gauge coupling universality,
$\delta g_{Wtb}$,
in a simple scenario where only the top quark Yukawa coupling is considered.
This occurs as the non-zero localized masses deviate mode functions.
Numerical calculation shows, however,
that the violation degree is not severe in mode parameter space.
Restricted in the KK mode degenerate case,
we have demonstrated quite significant correlation between
$\delta g_{Wtb}$ and the first KK mass of top quark
with respect to light quark KK mass.

\acknowledgments
This work was supported by the Korea Research Foundation
Grant KRF-2004-003-C00051.
S.C. is supported in part by Brainpool program of the KOFST.


%

\begin{thebibliography}{99}


\bibitem{dynamical}
S.~Weinberg,
Phys.\ Rev.\ D {\bf 13}, 974 (1976);
Phys.\ Rev.\ D {\bf 19}, 1277 (1979);
L.~Susskind,
Phys.\ Rev.\ D {\bf 20}, 2619 (1979).

\bibitem{Maldacena:1997re}
J.~M.~Maldacena,
Adv.\ Theor.\ Math.\ Phys.\  {\bf 2}, 231 (1998) [Int.\ J.\
Theor.\ Phys.\  {\bf 38}, 1113 (1999)] [arXiv:hep-th/9711200].
%
\bibitem{Gubser:1998bc}
S.~S.~Gubser, I.~R.~Klebanov and A.~M.~Polyakov,
Phys.\ Lett.\ B {\bf 428}, 105 (1998) [arXiv:hep-th/9802109].
%
\bibitem{Witten:1998qj}
E.~Witten,
Adv.\ Theor.\ Math.\ Phys.\  {\bf 2}, 253 (1998)
[arXiv:hep-th/9802150].

\bibitem{Aharony:1999ti}
O.~Aharony, S.~S.~Gubser, J.~M.~Maldacena, H.~Ooguri and Y.~Oz,
Phys.\ Rept.\  {\bf 323}, 183 (2000) [arXiv:hep-th/9905111].
%

\bibitem{Randall:1999ee}
L.~Randall and R.~Sundrum,
Phys.\ Rev.\ Lett.\  {\bf 83}, 3370 (1999) [arXiv:hep-ph/9905221].

\bibitem{Randall:1999vf}
L.~Randall and R.~Sundrum,
Phys.\ Rev.\ Lett.\  {\bf 83}, 4690 (1999) [arXiv:hep-th/9906064].


\bibitem{Arkani-Hamed:2000ds}
N.~Arkani-Hamed, M.~Porrati and L.~Randall,
JHEP {\bf 0108}, 017 (2001) [arXiv:hep-th/0012148].
%

\bibitem{Rattazzi:2000hs}
R.~Rattazzi and A.~Zaffaroni,
JHEP {\bf 0104}, 021 (2001) [arXiv:hep-th/0012248].
%

\bibitem{Perez-Victoria:2001pa}
M.~Perez-Victoria,
JHEP {\bf 0105}, 064 (2001) [arXiv:hep-th/0105048].

\bibitem{Davoudiasl:1999tf}
H.~Davoudiasl, J.~L.~Hewett and T.~G.~Rizzo,
Phys.\ Lett.\ B {\bf 473}, 43 (2000) [arXiv:hep-ph/9911262].

\bibitem{Chang:1999nh}
S.~Chang, J.~Hisano, H.~Nakano, N.~Okada and M.~Yamaguchi,
Phys.\ Rev.\ D {\bf 62}, 084025 (2000) [arXiv:hep-ph/9912498].

\bibitem{Kim:2002kk}
C.~S.~Kim, J.~D.~Kim and J.~Song,
Phys.\ Rev.\ D {\bf 67}, 015001 (2003)
[arXiv:hep-ph/0204002].


\bibitem{Grossman:1999ra}
Y.~Grossman and M.~Neubert,
Phys.\ Lett.\ B {\bf 474}, 361 (2000) [arXiv:hep-ph/9912408].

\bibitem{Gherghetta:2000qt}
T.~Gherghetta and A.~Pomarol,
Nucl.\ Phys.\ B {\bf 586}, 141 (2000) [arXiv:hep-ph/0003129].


\bibitem{Huber:2000fh}
S.~J.~Huber and Q.~Shafi,
Phys.\ Rev.\ D {\bf 63}, 045010 (2001) [arXiv:hep-ph/0005286],

\bibitem{Huber:2000ie}
S.~J.~Huber and Q.~Shafi,
Phys.\ Lett.\ B {\bf 498}, 256 (2001) [arXiv:hep-ph/0010195].





\bibitem{Pomarol:2000hp}
A.~Pomarol,
Phys.\ Rev.\ Lett.\  {\bf 85}, 4004 (2000) [arXiv:hep-ph/0005293].


\bibitem{Goldberger:2002hb}
W.~D.~Goldberger and I.~Z.~Rothstein,
Phys.\ Rev.\ D {\bf 68}, 125011 (2003) [arXiv:hep-th/0208060].

\bibitem{Agashe:2002bx}
K.~Agashe, A.~Delgado and R.~Sundrum,
Nucl.\ Phys.\ B {\bf 643}, 172 (2002) [arXiv:hep-ph/0206099].

\bibitem{Contino:2002kc}
R.~Contino, P.~Creminelli and E.~Trincherini,
JHEP {\bf 0210}, 029 (2002) [arXiv:hep-th/0208002].

\bibitem{Randall:2001gb}
L.~Randall and M.~D.~Schwartz,
JHEP {\bf 0111}, 003 (2001) [arXiv:hep-th/0108114].

\bibitem{Randall:2001gc}
L.~Randall and M.~D.~Schwartz,
Phys.\ Rev.\ Lett.\  {\bf 88}, 081801 (2002)
[arXiv:hep-th/0108115].


\bibitem{Choi:2002ps}
K.~W.~Choi and I.~W.~Kim,
Phys.\ Rev.\ D {\bf 67}, 045005 (2003) [arXiv:hep-th/0208071].

\bibitem{Agashe:2002pr}
K.~Agashe, A.~Delgado and R.~Sundrum,
Annals Phys.\  {\bf 304}, 145 (2003) [arXiv:hep-ph/0212028].

\bibitem{Goldberger:2002pc}
W.~D.~Goldberger, Y.~Nomura and D.~R.~Smith,
Phys.\ Rev.\ D {\bf 67}, 075021 (2003) [arXiv:hep-ph/0209158].

\bibitem{KChoi}
 K.~w.~Choi, H.~D.~Kim and I.~W.~Kim,
  JHEP {\bf 0303}, 034 (2003)
  [arXiv:hep-ph/0207013];
  K.~w.~Choi, H.~D.~Kim and I.~W.~Kim,
  JHEP {\bf 0211}, 033 (2002)
  [arXiv:hep-ph/0202257].





\bibitem{Csaki:2002gy}
C.~Csaki, J.~Erlich and J.~Terning,
%
Phys.\ Rev.\ D {\bf 66}, 064021 (2002) [arXiv:hep-ph/0203034].

\bibitem{Hewett:2002fe}
J.~L.~Hewett, F.~J.~Petriello and T.~G.~Rizzo,
%
JHEP {\bf 0209}, 030 (2002) [arXiv:hep-ph/0203091].

\bibitem{Burdman:2002gr}
G.~Burdman,
Phys.\ Rev.\ D {\bf 66}, 076003 (2002) [arXiv:hep-ph/0205329].






\bibitem{Agashe:2003zs}
K.~Agashe, A.~Delgado, M.~J.~May and R.~Sundrum,
JHEP {\bf 0308}, 050 (2003) [arXiv:hep-ph/0308036].

\bibitem{Burdman:2004rz}
  G.~Burdman,
  AIP Conf.\ Proc.\  {\bf 753}, 390 (2005)
  [arXiv:hep-ph/0409322].


\bibitem{Csaki:2003dt}
C.~Csaki, C.~Grojean, H.~Murayama, L.~Pilo and J.~Terning,
Phys.\ Rev.\ D {\bf 69}, 055006 (2004) [arXiv:hep-ph/0305237].


\bibitem{Csaki:2003zu}
C.~Csaki, C.~Grojean, L.~Pilo and J.~Terning,
Phys.\ Rev.\ Lett.\  {\bf 92}, 101802 (2004)
[arXiv:hep-ph/0308038].


\bibitem{Csaki:2003sh}
C.~Csaki, C.~Grojean, J.~Hubisz, Y.~Shirman and J.~Terning,
Phys.\ Rev.\ D {\bf 70}, 015012 (2004) [arXiv:hep-ph/0310355].




\bibitem{Nomura:2003du}
Y.~Nomura,
JHEP {\bf 0311}, 050 (2003) [arXiv:hep-ph/0309189].


\bibitem{Barbieri:2004qk}
R.~Barbieri, A.~Pomarol, R.~Rattazzi and A.~Strumia,
Nucl.\ Phys.\ B {\bf 703}, 127 (2004) [arXiv:hep-ph/0405040].



\bibitem{Barbieri:2003pr}
R.~Barbieri, A.~Pomarol and R.~Rattazzi,
Phys.\ Lett.\ B {\bf 591}, 141 (2004) [arXiv:hep-ph/0310285].





\bibitem{Foadi:2003xa}
R.~Foadi, S.~Gopalakrishna and C.~Schmidt,
JHEP {\bf 0403}, 042 (2004) [arXiv:hep-ph/0312324].

\bibitem{Davoudiasl:2003me}
H.~Davoudiasl, J.~L.~Hewett, B.~Lillie and T.~G.~Rizzo,
%
Phys.\ Rev.\ D {\bf 70}, 015006 (2004) [arXiv:hep-ph/0312193].


\bibitem{Davoudiasl:2004pw}
H.~Davoudiasl, J.~L.~Hewett, B.~Lillie and T.~G.~Rizzo,
JHEP {\bf 0405}, 015 (2004) [arXiv:hep-ph/0403300].


\bibitem{Burdman:2003ya}
G.~Burdman and Y.~Nomura,
Phys.\ Rev.\ D {\bf 69}, 115013 (2004) [arXiv:hep-ph/0312247].



\bibitem{Cacciapaglia:2004jz}
G.~Cacciapaglia, C.~Csaki, C.~Grojean and J.~Terning,
Phys.\ Rev.\ D {\bf 70}, 075014 (2004) [arXiv:hep-ph/0401160].



\bibitem{Cacciapaglia:2004rb}
G.~Cacciapaglia, C.~Csaki, C.~Grojean and J.~Terning,
arXiv:hep-ph/0409126.



\bibitem{Birkedal:2004au}
A.~Birkedal, K.~Matchev and M.~Perelstein,
arXiv:hep-ph/0412278.


\end{thebibliography}
\end{document}